\documentclass[graybox]{svmult}

\usepackage{mathptmx}        % selects Times Roman as basic font
\usepackage{makeidx}         % allows index generation
\usepackage{cite}
\usepackage{graphicx}        % standard LaTeX graphics tool
\usepackage{grffile}
\usepackage{epstopdf}        % when including figure files
\usepackage{amsmath}
\usepackage{multicol}        % used for the two-column index
\usepackage[bottom]{footmisc}% places footnotes at page bottom

\makeindex             % used for the subject index
\begin{document}

\title*{Parallelizing multiple precision Taylor series method for integrating the Lorenz system}

\author{I. Hristov$^{1,a}$, R. Hristova$^1$, S. Dimova$^1$, P. Armyanov$^1$, N. Shegunov$^1$, \\ I. Puzynin$^2$,  T. Puzynina$^2$, Z. Sharipov$^{2,b}$, Z. Tukhliev$^2$\\
\vspace{0.5cm}
\emph{$^1$ Sofia University, Faculty of Mathematics and Informatics, Bulgaria}\\
\emph{$^2$ JINR, Laboratory of Information Technologies, Dubna, Russia}\\
\vspace{0.5cm}
\emph{E-mails:  $^a$  ivanh@fmi.uni-sofia.bg \hspace{0.3cm} $^b$ zarif@jinr.ru}}
\authorrunning{I. Hristov, R. Hristova et al.}
\maketitle

\textbf{Keywords}: OpenMP, MPI, Multiple precision, Taylor series method, Lorenz system\\
\textbf{Mathematics Subject Classification}: 65L05, 65Y05\\

\abstract {A hybrid MPI+OpenMP strategy for parallelizing multiple precision Taylor series method is proposed, realized and tested.
To parallelize the algorithm we combine MPI and OpenMP parallel technologies together with GMP library (GNU miltiple precision libary)
and the tiny MPIGMP library.
The details of the parallelization are explained on the paradigmatic model of the Lorenz system.
We succeed to obtain a correct reference solution in the rather long time interval - [0,7000].
The solution is verified by comparing the results for 2700-th order Taylor series method and precision of $\sim$ 3374 decimal digits, and those with 2800-th order and precision of $\sim$ 3510 decimal digits.
With 192 CPU cores in Nestum cluster, Sofia, Bulgaria, the 2800-th order computation lasted  $\sim$ 145  hours with  speedup  $\sim$ 105.}

\section{Introduction}
\label{sec:2}
Computing mathematically reliable long-term trajectories of a chaotic dynamical system is not a trivial task due to the sensitive dependence on the initial conditions.
However, the advances in numerical methods and computer technologies in recent years allow us to overcome these difficulties  and give us new opportunities to explore the chaos.
A key work in this direction is the paper  of Shijun Liao \cite{Liao1}. He considers a new numerical procedure called  "Clean Numerical Simulation" (CNS) to obtain verified numerical solutions of chaotic dynamical systems. The procedure is based on  the multiple precision Taylor series method \cite{Jorba, Barrio1, Barrio2}.

The main concept in \cite{Liao1} is the critical predictable time $T_c$, which is a kind of  practical Lyapunov time.
$T_c$ is defined as the time for decoupling of two trajectories computed by two different numerical schemes.
The CNS works as follows.
First, estimates of  the required order of the  method  $N$ and the required precision
(the number of exact decimal digits $K$ of the floating point numbers) are obtained.
The estimate of $K$ is obtained by computing the $T_c-K$ dependence by means of the numerical solutions for fixed large enough $N$.  This estimate is in fact an estimate for the maximum Lyapunov exponent \cite{Wang}. The optimal order $N$ is estimated by computing the $T_c-N$ dependence by means of the  numerical solutions for fixed large enough $K$. This choice of $N$ ensures that the round-off error and the truncation error are of the same order. The solution is then computed with the estimated  $N$ and $K$ and after that one more computation with  higher $N$ and $K$ is performed for verification. A review of CNS and its important applications can be found in \cite{CNS}.

The first parallelization of CNS is reported in \cite{par1} and later improved in \cite{par2}. It is explained in   \cite{par1, par2} that a parallel reduction for computing the sums, which appear in the formulas of Taylor coefficients needs to be done. Of course, this is the crucial observation,  but no details of the parallelization of the algorithm are given there. A correct reference solution of the Lorenz system  for a time interval of a record length, namely [0,10000], obtained in about 9 days and 5 hours by using the pretty large computational resource of 1200 CPU cores, is given in \cite{Liao2}.

The goal of this work is to present in more details a simple and efficient hybrid MPI+OpenMP parallelization of the multiple precision Taylor series method, which allows to use arbitrarily  large  computational resource, if needed.  We  find that together with the parallel reduction there is additional, small but still important parallelism that could be carried out. In this work we use a moderate computational resource. With  192 CPU cores we compute a correct reference solution in the time interval [0,7000]. Also we have estimated the time  our program needs for computing the reference solution in the time interval [0,10000] with the same floating point precision, the step size and order of the method as in \cite{Liao2}. When using 256 CPU cores, the estimated time is $\sim$ 13 days and 12 hours with parallel efficiency $\sim$ 53\% and speedup $\sim$ 136.

Although our test model is the classical Lorenz system, the proposed parallelization strategy is rather general - it could be applied as well to a large class of chaotic dynamical systems.

\section{Taylor series method for the Lorenz system}
\label{sec:3}
We consider as a model problem the classical Lorenz system \cite{Lorenz}:
\begin{equation}
\begin{aligned}
\frac{dx}{dt} &= \sigma(y-x)\\
\frac{dy}{dt} &= Rx - y -xz\\
\frac{dz}{dt} &= xy - bz,
\end{aligned}
\end{equation}
where $R=28$, $\sigma=10$, $b=8/3$  are the standard Salztman's parameter values.
For these parameters the system is chaotic.
The N-th order Taylor series method \cite{Jorba, Barrio1, Barrio2} for (1) with step size $\tau$ is:
\begin{equation}
\begin{aligned}
x_{n+1} &= x_{n} + \sum_{i=1}^{N} \alpha_i \tau^i,\\
y_{n+1} &= y_{n} + \sum_{i=1}^{N} \beta_i \tau^i,\\
z_{n+1} &= z_{n} + \sum_{i=1}^{N} \gamma_i \tau^i,
\end{aligned}
\end{equation}
where
$$
\begin{aligned}
\alpha_i &= \frac{1}{i!}\frac{d^i x(t_n)}{{dt}^i},\\
\beta_i &= \frac{1}{i!}\frac{d^i y(t_n)}{{dt}^i}, \\
\gamma_i &= \frac{1}{i!}\frac{d^i z(t_n)}{{dt}^i}
\end{aligned}
$$
are the i-th Taylor coefficients (the so called  {\it{normalized derivatives}}).
They are computed as follows. From the system (1) we have
$$
\begin{aligned}
\alpha_1 &= \sigma(\beta_0 - \alpha_0),\\
\beta_1 &= R \alpha_0 - \beta_0 - \alpha_0 \gamma_0,\\
\gamma_1 &= \alpha_0 \beta_0 -b \gamma_0,
\end{aligned}
$$
where $$\alpha_0 = x_n, \hspace{0.5cm} \beta_0 = y_n, \hspace{0.5cm} \gamma_0 = z_n. $$

By applying Leibniz rule for the derivatives of the product of two functions,
we obtain the following recursive procedure for computing  $\alpha_i, \beta_i, \gamma_i$  for $i=0,..., N-1$:

\begin{equation}
\begin{aligned}
\alpha_{i+1}  &= \frac{1}{i+1} \sigma (\beta_i - \alpha_i),\\
\beta_{i+1}  &= \frac{1}{i+1} (R\alpha_i -\beta_i -\sum_{j=0}^{i}\alpha_{i-j}\gamma_j),\\
\gamma_{i+1}  &= \frac{1}{i+1} (\sum_{j=0}^{i}\alpha_{i-j}\beta_j -b\gamma_i).
\end{aligned}
\end{equation}

Note that we do not need any analytical expressions for the derivatives of $x(t)$, $y(t)$, $z(t)$.
 We only need the values of the derivatives at the point $t_n$.
Let us store the Taylor coefficients in the arrays \textbf{x}, \textbf{y}, \textbf{z} of lengths N+1. The values of $\alpha_i$ are stored in \textbf{x[i]},
those of $\beta_i$ in \textbf{y[i]} and those of $\gamma_i$ in \textbf{z[i]}. The pseudocode of the Taylor series method for the Lorenz system expressed by C-code in double precision is given in Figure 1. To compute the {\it{i+1}}-st coefficient in the Taylor series we need all previous coefficients from 0 to {\it{i}}. It is obvious  that we need $O(N^2)$ floating point operations for computing all coefficients. The subsequent evaluation of Taylor series with Horner's rule needs only $O(N)$ operations.

Actually, the  algorithm for computing the coefficients of the Taylor series, explained above,
is called {\it{automatic differentiation}}, or sometimes {\it{algorithmic differentiation}} \cite{Moore}.
Generally speaking, the automatic differentiation is a recursive procedure for computing the derivatives of certain functions
at a given point without using analytical formulas for the derivatives.
By "certain functions" we mean functions that can be obtained by sum, product,
quotient, and composition of some elementary functions. It is important that in all cases of dynamical systems whose right-hand side is
automatically differentiable,  sums like those in (3) are obtained. Thus, the approach for parallelization of our model problem
can be applied straightforwardly to a large class of dynamical systems.
\begin{figure}
\begin{center}

\includegraphics[scale=0.92]{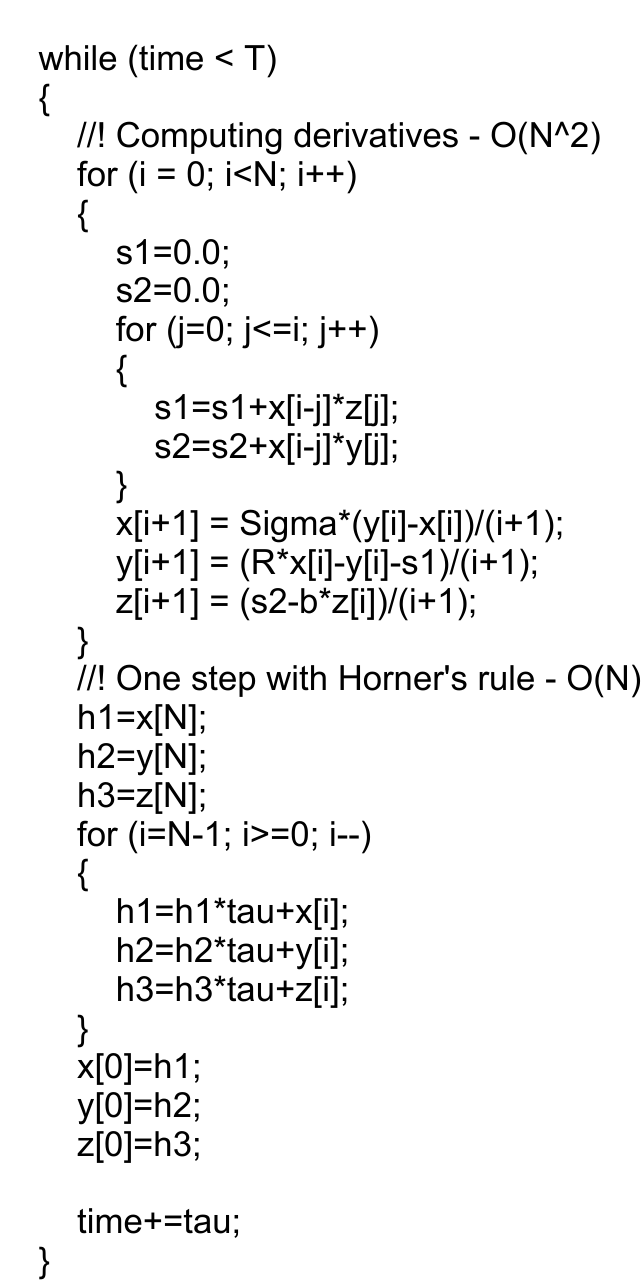}
\caption{Pseudocode of Taylor series method for the Lorenz system}
\label{fig:1}
\end{center}
\end{figure}

\section{Parallelization of the algorithm }

Let us look at the pseudocode in Figure 1. It is clear that the crucial decision for parallelization is to make a parallel reduction for the sums \textbf{s1} and \textbf{s2}, because this makes the algorithm complexity  $O(N^2)$. The same conclusion is found in \cite{par1, par2}. This parallelism however is not the only one  that could be utilized. Computing \textbf{x[i+1]}, \textbf{y[i+1]}, \textbf{z[i+1]} from formulas (3) can also be performed in parallel. We can also compute  the new values of \textbf{x[0]}, \textbf{y[0]}, \textbf{z[0]} in parallel by separating three independent Horner's loops. The last parallelism is of course a limited one, but by reducing the remaining serial part of the code (the part except reduction), we improve the Amdal's law. This additional parallelism is not perceptible for small number of cores, but plays its role in the case of large number of cores.

To parallelize the algorithm we combine MPI \cite{MPI} and OpenMP \cite{OpenMP,OpenMP2} parallel technologies together with GMP library (GNU miltiple precision libary)  \cite{gnu}.
We consider a hybrid MPI+OpenMP strategy, i.e. every MPI process creates a team of OpenMP threads. The main reason to consider a hybrid strategy, rather than a pure MPI strategy is that the results for speedup with OpenMP on one computational  node are slightly better than those with pure MPI.

We want to point out two additional  advantages of pure OpenMP over pure MPI for our problem.
We can not apply the domain decomposition for our problem,
because the algorithm is such that each MPI process needs access to each element of the arrays \textbf{x}, \textbf{y}, \textbf{z}.
So, the  memory needed for one computational node is multiplied by the number of the MPI processes per that node,
while OpenMP needs only one copy of the computational domain and thus some memory is saved.
Pure OpenMP is important also because it is easier to program. The reason is that the communication between threads is achieved by the shared memory and it is not necessary to take care of how to group GMP multiple precision data for explicit communications.
Using MPI and learning other libraries  for grouping GMP data for messages, however is unavoidable if we want to achieve a more massive parallelism.

The sketch of the hybrid MPI+OpenMP code corresponding to one step of Taylor method is shown in Figure 2.
The entire code for one step is contained in one OpenMP parallel region in order to maximize the region and to reduce the overall parallel overhead.
The parallel overhead for the OpenMP directives in the first nested loop however are multiplied by the order $N$ of the method and this is
unavoidable. As a consequence, the parallel scalability is expected to get worse for a fixed multiple precision and increasing order of the method,
and vice versa - to get better for a fixed order of the method and increasing precision.

The directive \textbf{\#pragma omp parallel private(i, j, tid)} creates a parallel region - additional threads, beside the master thread, are awaken.
The integer variables \textbf{i},  \textbf{j} and \textbf{tid} are defined as private in the parallel region (each thread has its own copy in its own stack).
First, every thread gets its \textbf{id} and stores it in \textbf{tid} by using the library function \textbf{omp\_get\_thread\_num()}.
Then, the  loop with index \textbf{i} is performed. Every MPI process takes its portion - the first and the last index controlled by the process.
The portions are computed in an OpenMP \textbf{single} section by standard formulas (\cite{quin}, Chapter 5).
After that the directive \textbf{\#pragma omp for} shares the work for
the loop between threads, i.e. every thread works on its own portion of the range [istart,...,ifinal].
The default static schedule clause is expected to work best, because the work is predictable and balanced for each loop index.
We use containers for the partial sums of every thread and these containers are shared. We store the containers in an array of multiple precision numbers \textbf{sum}.
We have in addition an array of temporary variables \textbf{tempv} for storing the intermediate results of multiplications.
To avoid false sharing, a padding strategy is applied for arrays \textbf{sum} and \textbf{tempv} \cite{OpenMP, OpenMP2}.
In fact, false sharing does not have the usual dramatic effect on the performance,
because the accompanying multiple precision operations are very time consuming.
The bracket that closes the parallel \textbf{for} loop acts as an implicit barrier for synchronization of the threads.

Although OpenMP has a build-in reduction clause,  we can not use it, because we use user-defined types for multiple precisions number and user-defined operations.
Thus, we have to do the reduction manually. We apply a standard tree based parallel reduction and the number of stages is only logarithm of the number of threads. Since we do the first
step in a butterfly form, we also ensure that only one multiple precision addition is done on every stage. The sums \textbf{s1} and \textbf{s2}
 are stored in \textbf{sum[0]} and \textbf{sum[1]} respectively.

At the point where each process has computed its partial sums, we are ready to perform  MPI\_ALLREDUCE \cite{MPI}.
To explain to MPI how to package and unpackage GMP multiple precision types, one needs good knowledge of both underlying representation of these types
and MPI. Tomonori Kouya done an excellent work by creating additional libraries for MPI programs which want to use  MPFR \cite{mpfr} and GMP multiple precision libraries.  We rely on the tiny  MPIGMP library of Tomonori Kouya \cite{Kouya0, Kouya1, Kouya2, Kouya3} for a straightforward usage of MPI\_ALLREDUCE.

The MPI\_ALLREDUCE is executed by the master threads of each process, but this is done inside the OpenMP parallel region.
It is reasonable  to look at the MPI\_ALLREDUCE as a continuation of the tree based reduction process, which starts with the OpenMP reduction.
Communications between master threads are overlapped with some computations for \textbf{x[i+1]}, \textbf{y[i+1]}, \textbf{z[i+1]} that can be taken in advance before the computation of the sums \textbf{s1} and \textbf{s2} is finished. One thread computes \textbf{x[i+1]}, other thread computes \textbf{Rx[i]-y[i]} from the formula for \textbf{y[i+1]}, and third thread computes \textbf{bz[i]} from the formula for \textbf{z[i+1]}.
When the MPI\_ALLREDUCE  is finished, we compute in parallel the rest operations for \textbf{y[i+1]}, \textbf{z[i+1]}. Since multiplication is much more expensive then the other used operations (division by an integer number is not so expensive), by hiding part of the computations behind  the reduction, we got some performance benefit.
At the end of the loop with index \textbf{i} we set in parallel the partial containers for each thread to zero.

At last, the loop for computing the new values of \textbf{x[0]}, \textbf{y[0]}, \textbf{z[0]}
by Horner's rule is separated in three independent loops for computing them in parallel. We use the
\textbf{sections} construct for this purpose. Computing  \textbf{x[i+1]}, \textbf{y[i+1]}, \textbf{z[i+1]} and \textbf{x[0]}, \textbf{y[0]}, \textbf{z[0]}
independently in parallel, and hiding some computations behind MPI communications generally improves Amdahl's low.
The C-code in terms of GMP library of our hybrid MPI+OpenMP program can be downloaded from \cite{radahpc}.

\begin{figure}
\begin{center}
\includegraphics[scale=1.0]{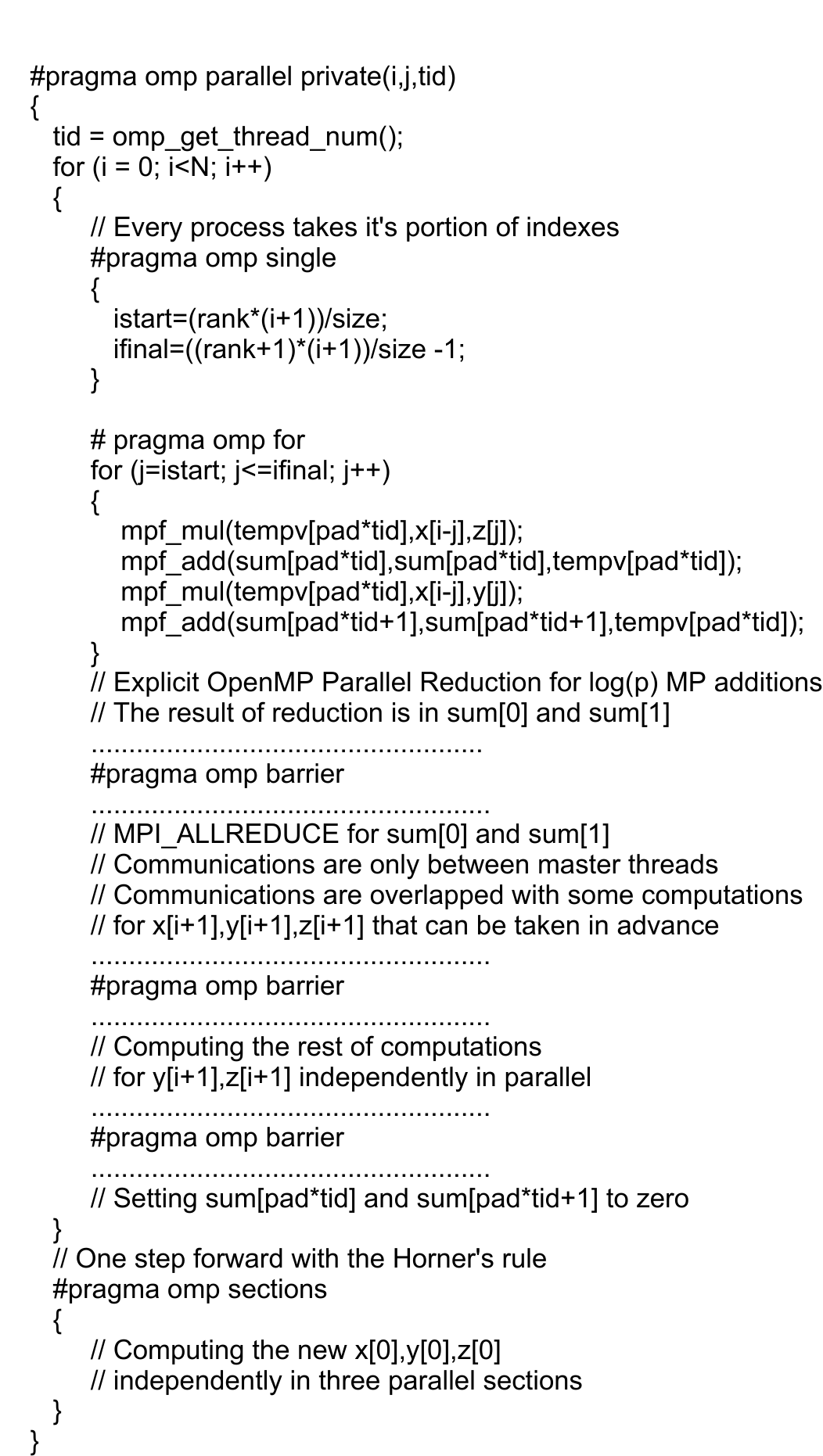}
\caption{The sketch of hybrid MPI+OpenMP code in terms of GMP library.}
\label{fig:2}
\end{center}
\end{figure}

Let us mention that if one half of the OpenMP threads compute the sum \textbf{s1} and the other half compute the sum \textbf{s2},
one could also expect some small performance benefit, because for the small indexes \textbf{i} unused threads
will be less and also the difference from the perfect load balance between threads will be less.
Thus we will have again a small improvement of Amdahl's low. However the last approach is not general because it strongly depends
on the number of sums for reduction (two in our case) and the number of available threads.

\section{Parallel performance scalability}

Part of computations are performed in the \textbf{HybriLIT} Heterogeneous Platform at the
Laboratory of IT of JINR, Dubna \cite{HybriLIT} and other part in the {\textbf{Nestum} Cluster, Sofia, Bulgaria \cite{nestum}.
The presented results for performance scalability are from Nestum Cluster. Nestum is a homogeneous HPC cluster based
on two socket nodes. Each node consists of 2 x Intel(R) Xeon(R) Processor E5-2698v3 (Haswell)  with 32 cores at 2.3 GHz.
We have used  Intel C++ compiler version 17.0, GMP library version 6.2.0, OpenMPI version 3.1.2 and compiler optimization options
-O3 -xhost.

As a benchmark we use the results for the Lorenz system with initial conditions and step size taken from \cite{Liao2}, namely $x(0)=-15.8$, $y(0)=-17.48$,
$z(0)=35.64$, $\tau=0.01$. The  computed $T_c-K$ dependence is $T_c \approx 2.5K$ and the computed $T_c-N$ dependence is  $T_c \approx 3N$ (\cite{Liao2}).
With a moderate computational resource, our goal was to obtain a correct reference solution for the time interval [0,7000] and to compare it with the results from Table 1 in
\cite{Liao2}. We carried out two  computations with 192 CPU cores (6 nodes in Nestum). We took the estimated values of $N$ and $K$ for $T_c=7000$ with some reserve, shown in brackets.
The first computation was with $N=2700$ $(\sim 15\%)$, $K \sim 3374$ $(\sim 20\%)$. The second computation was with $N=2800$ $(\sim 20\%)$, $K \sim 3510$ $(\sim 25\%)$.
The reserve is probably larger than necessary, but we wanted to be sure we got a correct reference solution with sufficient number of significant  digits.
All digits from \cite{Liao2} are the same as ours.
The reference solution for the interval [0,7000] with 30 correct digits can be downloaded from \cite{radahpc}.
We give below some strong scalability results for $N=2800$ and $K \sim 3510$.

As mentioned in the previous section, the main reason to use a hybrid strategy is that OpenMP scalability is slightly better than pure MPI inside one computational
node. For $N=2800$ and $K \sim 3510$, with one node (32 cores), we obtain  speedup $\sim 23.5$ and parallel efficiency of $ \sim 73.3 \%$ with OpenMP.
In comparison, we obtain speedup $\sim 22.3$ and parallel efficiency of  $ \sim 69.8 \%$ with pure MPI.
A possible explanation of the superior performance of OpenMP is that MPI has additional overhead for packaging and unpackaging the multiple precision types.

We also try  two MPI processes per node with one process for
each socket. This choice matches the node topology perfectly ( \cite{Hager}, Chapter 11).
For the above parameters however, we observe the same speedup for pure OpenMP and for one MPI process per socket.
Generally, the speedup results depend on the order of the method, the precision and the specific features of the HPC cluster used.
So, one should always try one MPI process per socket. Another  reason to do that is the fact that the memory access pattern of our algorithm is irregular and complicated.
One should think about possible NUMA-effects \cite{OpenMP,OpenMP2}, because every thread access every element from arrays \textbf{x, y, z}.
Existence of a thread from one socket which access memory from the other socket is unavoidable with one MPI process per node.

\begin{figure}
\begin{center}
\includegraphics[scale=0.55]{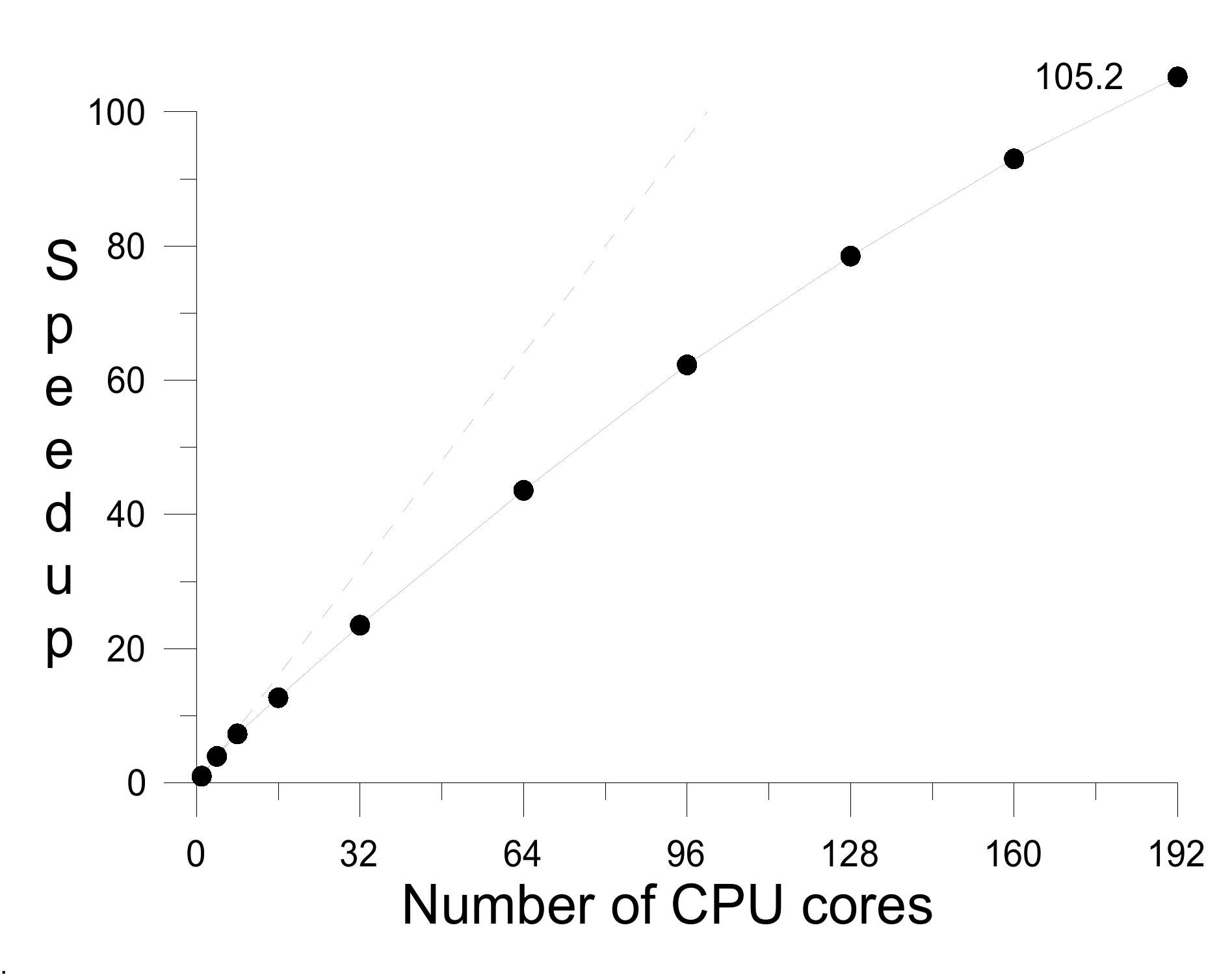}
\caption{Performance scalability for order of the method 2800 and precision 3510 decimal digits}
\label{fig:2}
\end{center}
\end{figure}

The results for the performance scalability for order of the method $N=2800$ and precision  $K \sim 3510$ decimal digits
obtained by the program sketched in Figure 2 are shown in Figure 3.
We use up to 192 CPU cores (6 nodes in Nestum cluster). The results for the performance scalability are pretty good.
The speedup using all 192 cores is about $\sim 105.2$, with parallel efficiency about $\sim 54.8 \%$. The correct reference solution for [0,7000] was computed with 192 CPU cores
within $\sim 145$ hours (6 days and 1 hour). Note, that it is always an important  idea to find a balance between using the resource efficiently and solving the problem
in a foreseeable time. For example, if we use only 64 cores (2 nodes), we will obtain the same result within the foreseeable time $\sim$ 14 days and 15 hours, but we will use the resource more efficiently.

In addition, we estimated the time needed for computing the reference solution for [0,10000] with $N=3500$, $K \sim 4180$, $\tau=0.01$, i.e. with the same parameters as
for the record simulation in \cite{Liao2}. The estimated time is $\sim$ 13 days and 12 hours using 256 CPU cores (8 nodes),
with parallel efficiency $\sim$ 53\% and speedup $\sim$ 136.

\section{Conclusions}
We proposed a  simple enough and efficient hybrid MPI+OpenMP parallelization of the multiple precision Taylor series method, which allows us to use arbitrarily large  computational resource, if needed. Although our test model is the classical Lorenz system, the proposed parallelization strategy is rather general, and it could be applied as well to a large class of chaotic dynamical systems in order to obtain long-term mathematically reliable solutions.

\begin{acknowledgement}
We thank the Laboratory of Information Technologies of JINR, Dubna, Russia for the opportunity to use the computational resources of the HybriLIT Heterogeneous Platform.
We also thank the opportunity to use the computational resources of the Nestum cluster, Sofia, Bulgaria.
We would like to give our special thanks to Stoyan Pisov for his great help in using the Nestum cluster.
The work is supported by a grant of the Plenipotentiary Representative of the Republic of Bulgaria at JINR, Dubna, Russia.
\end{acknowledgement}

\end{document}